\documentclass[aps,twocolumn,showpacs,floatfix]{revtex4-1}
\usepackage{dcolumn}
\usepackage{bm}
\usepackage{epsfig}
\usepackage{amsmath}
\usepackage{amssymb}
\usepackage{accents}
\usepackage{soul}

\usepackage{tikz}
\usepackage{graphicx}
\usetikzlibrary{arrows,shapes,trees,positioning,decorations.markings,decorations.pathmorphing}
\tikzset{
  ->-/.style={decoration={
  markings,
  mark=at position .5 with {\arrow{>}}},postaction={decorate}}
}
\tikzset{
    photon/.style={decorate, decoration={snake}, draw=red}}

\newcommand{\sbt}{\scalebox{.5}[.5]{\fboxsep=0ex\makebox[1ex][c]{$\bullet$}}}
\DeclareRobustCommand{\bull}[1]{\accentset{\sbt}{#1}}
\newcommand{\Di}{\Delta_\mathrm{i}}

\begin{document}
\title{Mutual Entropy-Production and Sensing in Bipartite Systems}
\author{Giovanni Diana}
\author{Massimiliano Esposito}
\affiliation{
Complex Systems and Statistical Mechanics, University of Luxembourg, L-1511 Luxembourg, Luxembourg
}

\begin{abstract}
We introduce and analyze the notion of mutual entropy-production (MEP) in autonomous systems. Evaluating MEP rates is in general a difficult task due to non-Markovian effects. For bipartite systems, we provide closed expressions in various limiting regimes which we verify using numerical simulations. Based on the study of a biochemical and an electronic sensing model, we suggest that the MEP rates provide a relevant measure of the accuracy of sensing. 
\end{abstract}

\maketitle

\section{Introduction}

A major achievement of the last decade, has been to establish a nonequilibrium thermodynamic description for small systems described by Markovian stochastic dynamics \cite{Sekimoto10, Seifert12Rev, JarzynskiRev11, Esposito12, QianPR12a, QianPR12b}. This theory, called stochastic thermodynamics, has close and remarkable connections with information theory. Various central quantities are expressed in terms of mathematical objects commonly used in information theory. The entropy of the system is given by its Shannon entropy \cite{Shannon48, CoverThomas} and the entropy production is a relative entropy between the probability of forward and time-reversed trajectories \cite{QianQian85, Crooks99, VandenBroeck07}. In a nonequilibrium steady-state, this latter can be expressed as a difference between the entropy rate \cite{CoverThomas, GaspardWang93} associated to forward and time-reversed trajectories \cite{Gaspard04b, Gaspard07exp, ParrondoEdgarPRE12}. 

Beside these formal connections to information theory, the framework of stochastic thermodynamics was used in the recent years, to revisit the finite-time aspects of various problems involving information processing which were originally proposed for reversible transformations. Such problems include for example the thermodynamic description of systems subjected to feedback \cite{SagawaUedaPRL08, SagawaUedaPRL09, SagawaUedaPRL10, Horowitz10, PekolaAverinPRB11, HorowitzParrondo11, HorowitzParrondo11NJP, SeifertAbreu11, SeifertAbreuPRL12, SagawaUeda12, JarzynskyPNAS12, SeifertBaratoEPL13, ParrondoHoroSag12, Rosinberg12, EspositoSchaller12, KunduPRE12, Leff} (which systematize the study of Maxwell demons and Szilard engines), Landauer's principle \cite{Leff, GaspardAndrieuxEPL08, EspoVdB_EPL_11, EspoDianaBagci13, GrangerKantzEPL13}, Bennet's reversible computing and kinetic proofreading \cite{Leff, Bennett73, Bennett79, AndrieuxPNAS08, EspositoJSM10, Tu_SpeedAccrNatPhys12, PigolottiPRL13, QianPR12b, LeiblerPNAS12, CrooksStill12}. Experimental verifications of these results have also been proposed \cite{Leigh, SagawaSano, LutzNat12, Snider12, Bechhoefer12}. 

One of the important new contribution has been to establish a systematic thermodynamic description of measurement, feedback and erasure for systems controlled by external time-dependent forces \cite{SagawaUedaPRL09, SagawaUedaPRL10, EspoVdB_EPL_11, SagawaUeda12, ParrondoHoroSag12, SeifertAbreuPRL12, EspoDianaBagci13}. The key quantity is the mutual information established as a result of the interaction between the system and the measurement apparatus. Unfortunately many systems performing some sensing tasks, such as electronic detectors \cite{EspositoCuetaraGaspard11, EspositoStrassSchall12, EspositoBulnessGasp13} and biochemical cell receptors \cite{MehtaSchwabPNAS12, SeifertBaratoPRE13}, operate as autonomous systems where these notions do not extend since the mutual information does not change in time. The thermodynamic treatment of such continuous measurement processes is thus harder to treat and no unified approach is currently available. Recent studies have considered the rate of mutual information defined at the trajectory level and found that it is not bounded by, nor related to, the entropy production \cite{SeifertBaratoPRE13, SeifertBarato13}.

In this paper, we introduce the concept of mutual entropy-production (MEP) and show that it can be expressed as the difference between the mutual informations associated to the forward and the time-reversed trajectories. This result can be put in parallel with the result obtained by Gaspard \cite{Gaspard04b} showing that for Markov processes the entropy production is not immediately related to the entropy rate but can be expressed as the difference between the entropy rate associated to the forward and the time-reversed trajectories. The MEP rate, as well as the mutual information rate, is in general difficult to calculate because coarse-graining introduces non-Markovian features \cite{ParrondoEdgarPRE12, SeifertBarato13}. For bipartite systems, we find limiting cases where the MEP rate can be obtained analytically and expressed in terms of quantities previously studied in Ref.~\cite{Esposito12} and easy to calculate. We finally suggest the MEP rate as a meaningful measure for the accuracy of sensing. We support this choice by considering two specific models, one describing biochemical signal transduction and the other the detection of single electron transfers in mesoscopic devices.

The plan of the paper is as follows. In section \ref{sec:one} we review the different ways to define entropy rates for general random processes. In section \ref{sec:MEP} we introduce the MEP and consider the limits where it can be obtained analytically. In section \ref{sec:APP} we consider two model systems describing sensing processes. We use them to numerically verify our theoretical predictions and to suggest the MEP rate as a thermodynamically meaningful measure of the sensing accuracy. Conclusions are drawn in section \ref{conc}.

\section{Entropy and entropy production rates} \label{sec:one}

We consider an arbitrary stationary random process of duration $t$ observed in discrete time with a time step $\tau$. Each realization of the process gives rise to a sequence $\mathbf{Z}_N = \lbrace z_1, z_2, \cdots, z_N \rbrace$ with a probability ${\cal P}(\mathbf{Z}_N)$ where $N=t/\tau$. 

The Shannon entropy in the space of all possible trajectories and its corresponding time-reversed entropy are defined as
\begin{align}
H(N,\tau) &\equiv-\langle\ln {\cal P}(\mathbf{Z}_N)\rangle,\label{eq:iterf}\\
H^R(N,\tau) &\equiv-\langle\ln {\cal P}(\mathbf{ Z}^R_N)\rangle,\label{eq:iterb}
\end{align}
where $\mathbf{Z}_N^R$ denotes the time-reversed sequence of $\mathbf{Z}_N$ and the averages $\langle \cdot \rangle$ are taken over the ensemble of sequences of length $N$.  

The entropy production is defined as the difference between the Shannon entropy and its corresponding time-reversed entropy
\begin{align} 
\Delta_\mathrm{i} S (N,\tau) \equiv H^R(N,\tau)-H(N,\tau) \geq 0. \label{eq:SidefbyFT}
\end{align}
It is always non-negative because, using (\ref{eq:iterf}) and (\ref{eq:iterb}), it can be expressed as a relative entropy (Kullback-Leibler divergence) between ${\cal P}(\mathbf{ Z}_N)$ and ${\cal P}(\mathbf{ Z}^R_N)$. It vanishes if and only if the probability of any trajectory coincides with the probability of its corresponding time-reversed trajectory, ${\cal P}(\mathbf{ Z}^R_N)={\cal P}(\mathbf{Z}_N)$, i.e. when the dynamics satisfies detailed balance. We note that this definition of entropy production is mathematical and its relation to the thermodynamic notion of entropy production is difficult to asses in general. For Markovian processes, this connection can be made explicitly \cite{Gaspard04b} (see also \cite{Crooks99, Maes03, Seifert05}). Nevertheless, following Refs.~\cite{VdBParrondoPRE08, ParrondoEdgarPRE12}, we will use (\ref{eq:SidefbyFT}) as the definition of entropy production for an arbitrary random processes. A closely related definition is also used for classical and quantum Hamiltonian systems \cite{VandenBroeck07, EspositoReview}. 

The Shannon entropy as well as the entropy production depends both on $N$ and $\tau$. We will now consider the rates associated to these quantities in the large $N$ limit and at fixed $\tau$. The entropy rates at fixed $\tau$ corresponding to (\ref{eq:iterf}) and (\ref{eq:iterb}) are defined as \cite{CoverThomas, GaspardWang93}
\begin{align}
\mathring{H}(\tau) \equiv \lim_{N\rightarrow\infty}\frac{H(N,\tau)}{N\tau}, \label{EntropyRatestau}\\
\mathring{H}^R(\tau) \equiv \lim_{N\rightarrow\infty}\frac{H^R(N,\tau)}{N\tau}. \label{EntropyRatestauR}
\end{align}
The corresponding entropy production rate at fixed $\tau$ is thus defined as
\begin{align}
\mathring{S}_{\mathrm{i}}(\tau) \equiv \lim_{N\rightarrow\infty}\frac{\Di S(N,\tau)}{N\tau}, \label{eq:sring}
\end{align}
and can be expressed in terms of the entropy rates (\ref{EntropyRatestau})-(\ref{EntropyRatestauR}) 
\begin{align}
\mathring{S}_{\mathrm{i}}(\tau)=\mathring{H}^R(\tau)-\mathring{H}(\tau).
\end{align} 

For a Markovian process described by a transition rate matrix $W_{z'z}$ (characterizing the probability per unit time to jump from a state $z$ to a state $z'$), the entropy rates (\ref{EntropyRatestau})-(\ref{EntropyRatestauR}) can be expressed as \cite{Gaspard04b}
\begin{align}
\mathring{H}(\tau)&=-\sum_{z\neq z'} W_{zz'}p(z') \ln W_{zz'} +B(\tau)\label{eq:hmarkov}\\ 
\mathring{H}^R(\tau)&=-\sum_{z\neq z'} W_{zz'}p(z') \ln W_{z'z} +B(\tau)\label{eq:hrevmarkov} 
\end{align} 
where $B(\tau) \equiv - \sum_{z\neq z'} W_{zz'}p(z') \ln \tau/e+\mathcal{O}(\tau)$. The symbol $\sum_{z \neq z'}$ denotes a summation over $z$ and $z'$ such that $z \neq z'$. The crucial observation made in Ref. \cite{Gaspard04b} is that while (\ref{eq:hmarkov}) and (\ref{eq:hrevmarkov}) depend on $\tau$ via $B(\tau)$, the entropy rate $\mathring{S}_\mathrm{i}$ does not and coincides with the well known entropy production rate \cite{Schnakenberg, NicoPrig77, Seifert05} 
\begin{align}
\sigma \equiv \sum_{z\neq z'} W_{zz'}p(z') \ln\frac{W_{zz'}p(z')}{W_{z'z}p(z)} \geq 0. \label{eq:Smarkov}
\end{align}

We now turn back to the general $N$ and $\tau$-dependent definition of Shannon entropy and entropy production (\ref{eq:iterf})-(\ref{eq:SidefbyFT}). Instead of considering rates associated to the large $N$ limit at fixed $\tau$, we now take the large $N$ limit at fixed duration of the process $t=N \tau$. This limiting procedure provides the continuous-time limit of the entropy production
\begin{align}\label{ContLimEP}
\Delta_\mathrm{i} S(t) \equiv \lim_{N\rightarrow \infty} \Delta_\mathrm{i} S(N,t/N). 
\end{align}
Taking the time derivative of $\Delta_\mathrm{i} S(t)$ is an alternative way compared to (\ref{eq:sring}) to define an entropy production rate 
\begin{align}\label{ContLimEPrate}
\bull{S}_{\mathrm{i}}(t)&\equiv\frac{d\Di S(t)}{dt} .
\end{align} 
In general this rate depends on $t$. Since $\Delta_\mathrm{i}S(t)\vert_{t=0}=0$ and assuming that $\bull{S}_{\mathrm{i}}(t)$ eventually reaches a constant asymptotic value, the short and long time limit of $\bull{S}_{\mathrm{i}}(t)$ can be expressed as
\begin{align}
\bull{S}_{\mathrm{i}}(0)&=\lim_{t\rightarrow0}\frac{\Di S_Z(t)}{t} \label{zerotimeMEP} \\
\bull{S}_{\mathrm{i}}(\infty)&=\lim_{t\rightarrow\infty}\frac{\Di S_Z(t)}{t} .
\end{align}
A remarkable feature of Markovian processes is that $\bull{S}_{\mathrm{i},Z}(t)$ becomes independent of $t$ and equal to (\ref{eq:Smarkov}). Therefore, for stationary Markov processes the rates (\ref{eq:sring}) and (\ref{ContLimEPrate}) coincide with (\ref{eq:Smarkov}), namely
\begin{equation} \label{eq:siudiff}
\bull{S}_{\mathrm{i}}=\mathring{S}_{\mathrm{i}}=\sigma_Z.
\end{equation}  
However, for general non-Markovian processes these entropy production rates do not necessarily coincide. 

\section{Mutual information and mutual entropy-production} \label{sec:MEP}

We now consider a Markovian random process $\mathbf{Z}=(\mathbf{X},\mathbf{Y})$ producing sequences in a space spanned by the pair of variables $z_i=(x_i,y_i)$. The random processes $\mathbf{X}$ and $\mathbf{Y}$ are a coarse-grained description of the joint Markovian process $\mathbf{Z}$ and are thus in general non-Markovian. 

A measure of the correlation between $\mathbf{X}$ and $\mathbf{Y}$ is the non-negative mutual information defined as \cite{CoverThomas}
\begin{eqnarray}
I(N,\tau) &\equiv& {H}_X(N,\tau) + {H}_Y(N,\tau) - {H}_{XY}(N,\tau) . \label{DefMutInf}
\end{eqnarray}
In analogy to what was done in the preceeding section for the Shanon entropy, we introduce the time-reversed mutual information as
\begin{eqnarray}
I^R(N,\tau) &\equiv& {H}^R_X(N,\tau)+{H}^R_Y(N,\tau)-{H}^R_{XY}(N,\tau).\label{DefMutInfR}
\end{eqnarray}
Pushing the analogy further, we introduce the concept of mutual entropy-production (MEP) which measures the difference between the entropy production of the joint process $\Di S_{XY}$ and the entropy production of the marginal processes $\Di S_{X}$ and $\Di S_{Y}$
\begin{align}
\Di S^M(N,\tau) \equiv \Di S_{XY}(N,\tau) - \Di S_{X}(N,\tau) - \Di S_{Y}(N,\tau). \label{eq:M}
\end{align}
Using (\ref{DefMutInf}) and (\ref{DefMutInfR}) with (\ref{eq:SidefbyFT}), we find that the MEP can be expressed as the difference between the mutual information and its time-reversal form
\begin{align}
\Di S^M(N,\tau)=I(N,\tau)-I^R(N,\tau) \label{eq:result1} .
\end{align}
This result is reminiscent of (\ref{eq:SidefbyFT}) and provides a connection between mutual information and entropy production. 

\subsection{Bipartite networks} \label{sec:BPnet}

To proceed with our analysis, we assume that the joint Markov process occurs on a bipartite network, where each transition between states $z$ can involve either a jump in $x$ or in $y$ but not in both. The transition matrix of the joint process is thus of the form 
\begin{equation}\label{eq:TM}
W^{x x'}_{y y'}\equiv\left\lbrace
\begin{array}{c c}
1- R^{x}_y \tau \quad & \mathrm{if}\; y=y'\; \mathrm{and}\; x= x'\\
w^{xx'}_y\tau\quad & \mathrm{if}\; y=y'\; \mathrm{and}\; x\neq x'\\
w^x_{yy'}\tau\quad  &\mathrm{if}\; x=x'\; \mathrm{and}\; y\neq y'\\
0 &\mathrm{if}\; x\neq x'\; \mathrm{and}\; y\neq y'
\end{array}
\right. ,
\end{equation}
where $R^{x}_y \equiv \sum_{x' (\neq x)} w^{x'x}_y + \sum_{y' (\neq y)} w_{y'y}^x$ is the decay rate from state $(x,y)$.
This transition matrix satisfies the normalization condition $\sum_{x,y} W^{xx'}_{yy'}=1$. 

In the continuous-time limit, the probability to find a system described by the transition matrix (\ref{eq:TM}) in a given state $(x,y)$ satisfies the Markovian master equation 
\begin{align}
\frac{d}{dt} p(x,y) = \sum_{x'} J_{xx'}(y) + \sum_{y'} J_{yy'}(x) ,\label{MasterEqJoint}
\end{align} 
where 
\begin{align}
& J_{xx'}(y) \equiv w^{xx'}_{y} p(x',y) - w^{x'x}_{y} p(x,y),  \label{ProbFlux} \\
& J_{yy'}(x) \equiv w_{yy'}^{x} p(x,y') - w_{y'y}^{x} p(x,y) \nonumber .
\end{align}  

Since the joint process $\mathbf{Z}$ is Markovian, the different definitions of the entropy production rate all coincide
\begin{align}
\bull{S}_{\mathrm{i},XY}=\mathring{S}_{\mathrm{i},XY}=\sigma_{XY} , 
\end{align} 
where
\begin{align}
\sigma_{XY} &=\sum_{y,x\neq x'} w^{x x'}_y  p(x',y)\ln \frac{w^{xx'}_y p(x',y)}{w^{x'x}_y p(x,y)} \nonumber\\
&+\sum_{x,y\neq y'} w^{x}_{yy'} p(x,y')\ln \frac{w^{x}_{yy'}p(x,y')}{w^x_{y'y}p(x,y)}. \label{EPSchnakenSep}
\end{align}
If the stochastic network contains multiple edges between pairs of nodes, the summations over pair of states in (\ref{EPSchnakenSep}) must contain a summation over all these edges. In other words, if the net transition rate between two states is in fact the sum of rates associated to different physical mechanisms $\nu$ such as reservoirs or chemical reactions (e.g. $w^{x'x}_y=\sum_{\nu} w^{x'x}_y (\nu)$), the summation in (\ref{EPSchnakenSep}) has to also contain the sum over $\nu$ \cite{EspoVdB10_Da, EspoVdB10_Db}.
 
Since the random processes $\mathbf{X}$ and $\mathbf{Y}$ constitute a coarse-grained description of the joint process $\mathbf{Z}$, they are in general non-Markovian. As a result, the MEP rate defined using the limiting procedure (\ref{ContLimEPrate}),  
\begin{align}
\bull{S}^M_\mathrm{i}(t) = \sigma_{XY} - \bull{S}_{\mathrm{i},X}(t) - \bull{S}_{\mathrm{i},Y}(t), \label{RateMEPfull}
\end{align}
does not necessarily coincide with the rate defined using (\ref{eq:sring}),
\begin{align}
\mathring{S}^M_\mathrm{i}(\tau) = \sigma_{XY} - \mathring{S}_{\mathrm{i},X}(\tau) - \mathring{S}_{\mathrm{i},Y}(\tau). \label{RateMEPempty}
\end{align}

\subsection{Decomposition of the entropy production} \label{sec:cgep}

Even though the MEP rates are often difficult to evaluate, we will see in section \ref{SecTzero} and \ref{sec:TSsep} that under special conditions they can be expressed in terms of much simpler quantities which appear in the general decomposition of the joint entropy production proposed in Ref.~\cite{Esposito12}. For bipartite networks this decomposition reads
\begin{align}
\sigma_{XY}=\sigma^{(1)}_X+\sigma^{(2)}_X+\sigma^{(3)}_X \geq 0 \label{eq:cgXslow},
\end{align} 
where
\begin{align}
\sigma_X^{(1)} \equiv &\sum_{x \neq x'}\overline{w}^{xx'}p(x') \ln \frac{\overline{w}^{xx'}p(x')}{\overline{w}^{x'x}p(x)} \geq 0 \label{eq:S1}\\
\sigma_X^{(2)} \equiv &\sum_{x, y \neq y'} w^{x}_{yy'} p(x,y') \ln \frac{w^{x}_{yy'}p(y'|x)}{w^{x}_{y'y} p(y|x)} \geq 0 \label{eq:S2}\\
\sigma_X^{(3)} \equiv &\sum_{x \neq x'} \overline{w}^{xx'}p(x') \sum_y f_y^{xx'}\ln\frac{f_y^{xx'}}{f_y^{x'x}} \geq 0 \label{eq:S3} .
\end{align} 
In these definitions we introduced the coarse-grained rates between $x'$ and $x$
\begin{align}
\overline{w}^{xx'} \equiv \sum_y w^{xx'}_y p(y|x'), \label{eq:effX}
\end{align}
where $p(y|x)$ is the conditional probability of finding $y$ given $x$, as well as 
\begin{align}
f_y^{xx'} \equiv \frac{w_y^{xx'}}{\overline{w}^{xx'}}p(y|x'), \label{Ffactors}
\end{align} 
the fraction of jumps between $x'$ and $x$ occurring at a given value of $y$, which is normalized by $\sum_{y} f_y^{xx'}=1$. 

The decomposition (\ref{eq:cgXslow}) is particularly useful when considering a description of the system in terms of the variable $x$ whereas $y$ has been coarse-grained. Indeed, the term $\sigma_X^{(1)}$ can be seen as an effective entropy production rate at the coarse-grained level and $\sigma_X^{(2)}$ as an average over $x$ of the various entropy productions due to the dynamics in $y$ at a given $x$. The last term $\sigma_X^{(3)}$ quantifies the asymmetry between $f_y^{xx'}$, the fraction of jumps occurring at a given $y$ between $x'$ and $x$,  and $f_y^{x'x}$, the fraction of jumps occurring at the same $y$ between $x$ and $x'$. This quantity is thus large when most of the transitions between $x'$ and $x$ occur at a given value of $y$ while most of the transitions between $x$ and $x'$ occur at a different value of $y$.

Analogously to (\ref{eq:cgXslow}), by exchanging the roles of $X$ and $Y$, we obtain the symmetric decomposition
\begin{align}
\sigma_{XY}=\sigma_Y^{(1)}+\sigma_Y^{(2)}+\sigma_Y^{(3)}, \label{eq:cgYslow}
\end{align} 
which is more relevant when the coarse-grained variable is $x$ instead of $y$.

By comparing (\ref{EPSchnakenSep}) with the definition (\ref{eq:S2}), we find that the Markovian entropy production rate for bipartite networks can be expressed as the sum
\begin{align}
\sigma_{XY}=\sigma_X^{(2)}+\sigma_Y^{(2)}.\label{EPBipNetSpecial}
\end{align} 
This property implies the following useful identities
\begin{align}
\sigma_X^{(2)}=\sigma_Y^{(1)}+\sigma_Y^{(3)} \ \ , \ \ \sigma_Y^{(2)}=\sigma_X^{(1)}+\sigma_X^{(3)}. \label{EPBipNetSpecialBis}
\end{align} 


\subsection{Short-time limit of the rates} \label{SecTzero}

To obtain an exact analytical expression for the MEP (\ref{RateMEPfull}), we will consider in this section its short-time limit
\begin{align}
\bull{S}^M_\mathrm{i}(0) = \lim_{t\rightarrow 0}\frac{\Di S^M(t)}{t} 
= \sigma_{XY} - \bull{S}_{\mathrm{i},X}(0) - \bull{S}_{\mathrm{i},Y}(0).
\end{align}
We start with the continuous-time limit of the MEP (\ref{ContLimEP}) which can be expressed as
\begin{align} \label{EPMuttraj}
&\Di S^M(t) = \sum_{\mathbf{Z}} \mathcal{P} \left( \ln \frac{\mathcal{P}}{\mathcal{P}^R}
-\ln \frac{\sum_{\mathbf{X}} \mathcal{P} }{\sum_{\mathbf{X}}\mathcal{P}^R}
-\ln \frac{\sum_{\mathbf{Y}} \mathcal{P} }{\sum_{\mathbf{Y}}\mathcal{P}^R}\right),
\end{align}
where the probabilities $\mathcal{P} \equiv \mathcal{P}(\mathbf{Z})$ and $\mathcal{P}^R \equiv \mathcal{P}(\mathbf{Z}^R)$ of the trajectories $\mathbf{Z}=(\mathbf{X},\mathbf{Y})$ are the continuous-time analogue of the discrete-time probabilities used in section \ref{sec:one}.

If we denote by $(\mathbf{X}^l,\mathbf{Y}^m)$ a trajectory with $l$ transitions in $X$ and $m$ transitions in $Y$, and if the initial state $(x,y)$ is drawn from the stationary probability $p(x,y)$, we get 
\begin{align}
\mathcal{P}(\mathbf{X}^0,\mathbf{Y}^0)&=p(x,y)(1-t R^{x}_y)\nonumber\\
\mathcal{P}(\mathbf{X}^1,\mathbf{Y}^0)&=p(x,y)t w^{x'x}_y\nonumber\\
\mathcal{P}(\mathbf{X}^0,\mathbf{Y}^1)&=p(x,y)t w^{x}_{y'y}\nonumber\\
\mathcal{P}(\mathbf{X}^l,\mathbf{Y}^m)&={\cal O}(t^2) \ \ {\rm if} \ \ l+m>1 . \label{eq:probs}
\end{align}
Using these expressions into (\ref{EPMuttraj}), we find that
\begin{align}
& \Di S^M (t) =\nonumber\\
& \sum_{x,x',y} tw^{x'x}_yp(x,y)\left(\ln\frac{w^{x'x}_yp(x,y)}{w^{xx'}_yp(x',y)}-\ln\frac{ \overline w^{x'x}p(x)}{\overline w^{xx'}p(x')}\right)\nonumber\\
&+ \sum_{x,y,y'}  tw^{x}_{y'y}p(x,y)\left(\ln\frac{w^{x}_{y'y}p(x,y)}{w^{x}_{yy'}p(x,y')}-\ln\frac{ \overline w^{y'y}p(y)}{\overline w^{yy'}p(y')}\right)\nonumber\\
&+\mathcal{O}(t^2), \label{eq:ctmepr}
\end{align}
which using (\ref{zerotimeMEP}) and (\ref{eq:S1}) leads to 
\begin{align}
\bull{S}^M_\mathrm{i}(0) = \sigma_{XY} - \sigma^{(1)}_X - \sigma^{(1)}_Y  \label{eq:mep0}.
\end{align}

The rate of MEP in the short-time limit is thus given by the entropy production of the joint system minus the sum of the effective entropy production resulting respectively from a coarse-graining over $x$ and $y$. Using the decomposition (\ref{eq:cgXslow}) and the relations (\ref{EPBipNetSpecialBis}), this result can also be rewritten as 
\begin{align}
\bull{S}^M_\mathrm{i}(0) = \sigma^{(3)}_X + \sigma^{(3)}_Y \geq 0 \label{eq:mep0Ineql}.
\end{align}
This important result shows that the short-time limit of the MEP rate does not depend explicitly on the terms $\sigma^{(1)}$ and $\sigma^{(2)}$, which characterize the dissipation along a given coordinate of the bipartite network. Instead, it can be exclusively expressed in terms of the $\sigma^{(3)}$, which characterize an intrinsically mixed source of dissipation.

We now turn to the rate of mutual information in the short-time limit and establish a connection with the work presented in Ref.~\cite{SeifertBarato13, SeifertBaratoPRE13}. 
The mutual information (\ref{DefMutInf}) in continuous-time can be expressed as 
\begin{align} \label{Itraj}
& I(t) = \sum_{\mathbf{Z}} \mathcal{P} \left( \ln \mathcal{P}-\ln \sum_{\mathbf{X}} \mathcal{P} 
-\ln \sum_{\mathbf{Y}} \mathcal{P} \right).
\end{align}
Using the short-time probabilities (\ref{eq:probs}), all terms proportional to $\ln t$ cancel out and only the constant and linear terms in $t$ survive. We thus obtain 
\begin{align}
I(t) =& M + t\sum_{x'xy} p(x,y)w^{x'x}\ln\frac{w^{x'x}}{\overline{w}^{x'x}}+\nonumber\\
&+t\sum_{y'yx}p(x,y)w^{x}_{y'y}\ln\frac{w^x_{y'y}}{\overline{w}_{y'y}}+\mathcal{O}(t^2),
\end{align} 
where $M$ is the mutual information associated to the steady-state probabilities
\begin{align}
M \equiv \sum_{x,y}p(x,y)\ln\frac{p(x,y)}{p(x)p(y)} \geq 0. \label{eq:Mxy}
\end{align}
The mutual information rate in the short-time limit is therefore given by
\begin{align}
&\bull{I}(0) = \frac{dI(t)}{dt}|_{t=0} =   \label{eq:ub}\\
&\sum_{xy} p(x,y) \bigg(\sum_{x'(\neq x)} w^{x'x}_y \ln\frac{w^{x'x}_y}{\overline{w}^{x'x}}
+\sum_{y'(\neq y)} w^{x}_{y'y}\ln\frac{w^x_{y'y}}{\overline{w}_{y'y}} \bigg) . \nonumber
\end{align}
We note that this quantity corresponds precisely to the upper bound of the rate of mutual information $\mathring{I}(\tau)$ in the limit $\tau \rightarrow 0$ found in Ref. \cite{SeifertBaratoPRE13,SeifertBarato13}, namely 
\begin{align}
\bull{I}(0) \geq \mathring{I}(0) . \label{conntoupperb}
\end{align}

Analogously, the time-reversed mutual information rate $\bull{I}^R(0)$ reads
\begin{align}
&\bull{I}^R(0) = \frac{dI^R(t)}{dt}|_{t=0}  =  \label{eq:ubR}\\
&\sum_{xy} p(x,y) \bigg(\sum_{x'(\neq x)} w^{x'x}_y \ln\frac{w^{xx'}_y}{\overline{w}^{xx'}}
+\sum_{y'(\neq y)} w^{x}_{y'y}\ln\frac{w^x_{yy'}}{\overline{w}_{yy'}} \bigg),  \nonumber
\end{align}
therefore we can also express the mutual information rate in the short-time limit as 
\begin{align}
\bull{S}^M_\mathrm{i}(0) = \bull{I}(0) - \bull{I}^R(0) \geq 0 .
\end{align}

\subsection{Time-scale separation}\label{sec:TSsep} 

A regime of time-scale separation occurs whenever the transitions in one of the two variables $X$ or $Y$ happen at a much higher rate than the other. In this section, we will assume that $Y$ is faster than $X$, thus the rates $w^x_{y'y}$ are much larger than $w^{xx'}_{y}$. To discuss this regime, we multiply the rates $w^{xx'}_{y}$ by a scaling factor $\gamma$. 
As shown in Ref.~\cite{Esposito12}, the marginal probability $p(x)$ satisfies always a master equation of the form
\begin{align}
\frac{d}{dt}p(x) = \sum_{x'} \left( \overline{w}^{xx'} p(x')-\overline{w}^{x'x} p(x) \right) , \label{EffMEq}
\end{align}
in terms of the effective rates $\overline{w}^{xx'}$ introduced in (\ref{eq:effX}). This equation is not closed since the effective rates depend on the conditional probabilities $p(y|x')$ which require the solution of the full joint dynamics (\ref{MasterEqJoint}). However, in the regime of time-scale separation, these probabilities can be obtained by finding the stationary state of the closed Markovian master equation (valid when $\gamma \to 0$)
\begin{align} 
\frac{d}{dt} p(y|x) = \sum_{y'} \left( w^{x}_{yy'} p(y'|x)- w^{x}_{y'y} p(y|x) \right) , \label{LocIntDyn}
\end{align} 
and used to calculate the effective rates (\ref{eq:effX}) perturbatively to order $\gamma$ \cite{Esposito12}. As a result, (\ref{EffMEq}) becomes a closed Markovian master equation and the entropy production rate for $X$ is thus given by
\begin{align} 
\bull{S}_{\mathrm{i},X} = \sigma^{(1)}_{X} + {\cal O}(\gamma^2) \label{eq:slowX}.
\end{align} 
The MEP rate (\ref{RateMEPfull}) therefore reduces to 
\begin{align}
\bull{S}^M_\mathrm{i}(t) &= \sigma_{XY} - \sigma^{(1)}_{X} - \bull{S}_{\mathrm{i},Y}(t) + {\cal O}(\gamma^2) \label{RateMEPfullTSS} \\
&= \sigma^{(3)}_X + \sigma^{(2)}_X - \bull{S}_{\mathrm{i},Y}(t)+ {\cal O}(\gamma^2), \nonumber 
\end{align}

To proceed, we consider the special case where the fast transitions between $y$ states do not depend on the states $x$. Using (\ref{eq:effX}) and (\ref{Ffactors}), we find that $\overline{w}^{xx'}=w^{xx'}$ and $f_y^{xx'}=p(y|x')$. Also, the conditional probabilities in (\ref{LocIntDyn}) become independent of $x$, i.e. $p(y|x)=p(y)$. The dynamics for $Y$ thus also becomes Markovian and
\begin{align} 
\bull{S}_{\mathrm{i},Y}=\sigma^{(1)}_Y. \label{MarkovY}
\end{align}
By inserting {(\ref{MarkovY})} into ({\ref{RateMEPfullTSS}}), we find that $\bull{S}^M_\mathrm{i}$ is independent of $t$ and, consistently with ({\ref{eq:mep0}}), it coincides with the short-time limit $\bull{S}^M_\mathrm{i}(0)$, also expressed as ({\ref{eq:mep0Ineql}}). Since the transition rates in $x$ do not depend on $y$, we also note from the definitions ({\ref{eq:S1}}) and ({\ref{eq:S2}}) that $\sigma^{(1)}_Y$ and $\sigma^{(2)}_X$ are equal. Thus the relations ({\ref{EPBipNetSpecialBis}}) imply that $\sigma^{(3)}_Y=0$. As a result, in this particular case the MEP rate reduces to
\begin{align} 
\bull{S}^M_\mathrm{i} = \sigma^{(3)}_X + {\cal O}(\gamma^2). \label{eq:tssres} 
\end{align}

We now turn to the situation where the fast process $\mathbf{Y}$ at fixed $x$ is locally at equilibrium for all $x$ in the limit of $\gamma \rightarrow 0$, i.e the conditional probabilities $p(y|x)$ satisfy the detailed balance relation
\begin{align} 
w^{x}_{y'y} p(y|x) = w^{x}_{yy'} p(y'|x) . \label{EquiInternalX}
\end{align}
As $\gamma \rightarrow 0$, $\sigma^{(2)}_X$ is of order $\gamma^2$, therefore from the relations (\ref{EPBipNetSpecialBis}) also $\sigma_Y^{(1)}$ and $\sigma_Y^{(3)}$ must be of the same order
\begin{align}
\sigma_X^{(2)}=\sigma_Y^{(1)}=\sigma_Y^{(3)}= {\cal O}(\gamma^2). \label{Ineq1-2}
\end{align} 
If we consider times $t$ shorter than the typical time needed for transitions between $x$ states to occur, i.e. $t \ll 1/\overline{w}_{xx'}\sim 1/\gamma$, the states $x$ are frozen and 
\begin{align}
{\Di S}_{Y}(t) = \left \langle \ln \frac{\sum_x p(x) \mathcal{P}(\mathbf{Y}|x)}{\sum_x p(x) \mathcal{P}(\mathbf{Y}^R|x)}\right\rangle_\mathbf{Y} , \label{IntEqSY}
\end{align}
where $p(x)$ is the probability to sample a trajectory starting (and thus staying) in $x$. 
Using the log-sum rule on (\ref{IntEqSY}), we find that $\bull{S}_{\mathrm{i},Y}(t) \leq \sigma_X^{(2)}$, which using (\ref{Ineq1-2}) and (\ref{RateMEPfullTSS}) implies
\begin{align} 
\bull{S}^M_\mathrm{i}(t) = \sigma^{(3)}_X+\mathcal{O}(\gamma^2)  \ \ {\rm for} \ \ t \ll 1/\overline{w}_{xx'} \sim 1/\gamma . \label{eq:tssresBis} 
\end{align}
This result is consistent with (\ref{eq:mep0Ineql}) as can be verified using (\ref{Ineq1-2}). 

However, for generic regimes of time-scale separation we have that 
\begin{align} \label{eq:SYs2text}
\bull{S}_{\mathrm{i},Y}(t) = \sigma^{(2)}_X + {\cal O}(\gamma) .
\end{align} 
Therefore, since
\begin{align}
\sigma^{(1)}_{X}, \sigma^{(3)}_{X}, \sigma^{(2)}_{Y} = {\cal O}(\gamma) \ \ , \ \
\sigma^{(1)}_{Y}, \sigma^{(3)}_{Y}, \sigma^{(2)}_{X} = {\cal O}(1) ,
\end{align}
the evaluation of the MEP rate (\ref{RateMEPfullTSS}) crucially depends on the corrections to (\ref{eq:SYs2text}) which are in general difficult to compute. 


\section{Applications} \label{sec:APP}

\subsection{Model system and numerical verifications} \label{ModelGen}

\begin{figure}[t]
\includegraphics{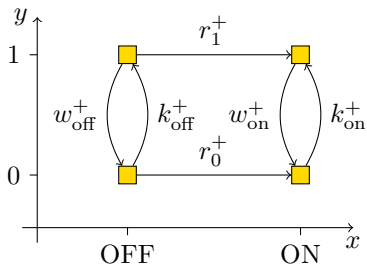}
\caption{\label{fig:model} Four-state model of a bipartite system made of two states $x={\rm off},{\rm on}$ and two states $y=0,1$. Each directed transition is associated to a rate with a superscript "$+$". The reversed transition is associated to the corresponding rate with superscript "$-$" (not displayed).}
\end{figure} 
In order to verify the results of section \ref{sec:TSsep}, we consider the bipartite model depicted in Fig.~(\ref{fig:model}) in the steady-state regime.
We also impose the condition
\begin{align}
a \equiv k^+_\mathrm{on}/k^+_\mathrm{off} = k^-_\mathrm{on}/k^-_\mathrm{off} .\label{SimpCond}
\end{align}
The Markovian entropy production of the joint system reads \cite{Schnakenberg}
\begin{align}
\sigma = J_\mathrm{c} \ln \frac{r_0^+ r_1^- }{r_0^- r_1^+} 
+ J_\mathrm{on} \ln \frac{k_{\rm on}^+ w^+_{\rm on}}{k_{\rm on}^- w^-_{\rm on}} 
+ J_\mathrm{off} \ln \frac{k_{\rm on}^+ w^+_{\rm off}}{k_{\rm on}^- w^-_{\rm off}} , \label{EPAff}
\end{align}
where 
\begin{align}
&J_\mathrm{c} = r_0^+ p({\rm off},0) - r_0^- p({\rm on},0) \nonumber\\
&J_\mathrm{on} = w^+_{\rm on} p({\rm on},1) - w^-_{\rm on} p({\rm on},0) \nonumber\\
&J_\mathrm{off} = w^+_{\rm off} p({\rm off},1) - w^-_{\rm off} p({\rm off},0) 
\end{align}
are respectively the counterclockwise probability currents associated to the large and the two small cycles on Fig.~(\ref{fig:model}). Equilibrium requires the three affinities (i.e. the logarithms in (\ref{EPAff})) to vanish.

\begin{figure}[h]
\includegraphics[scale=.7]{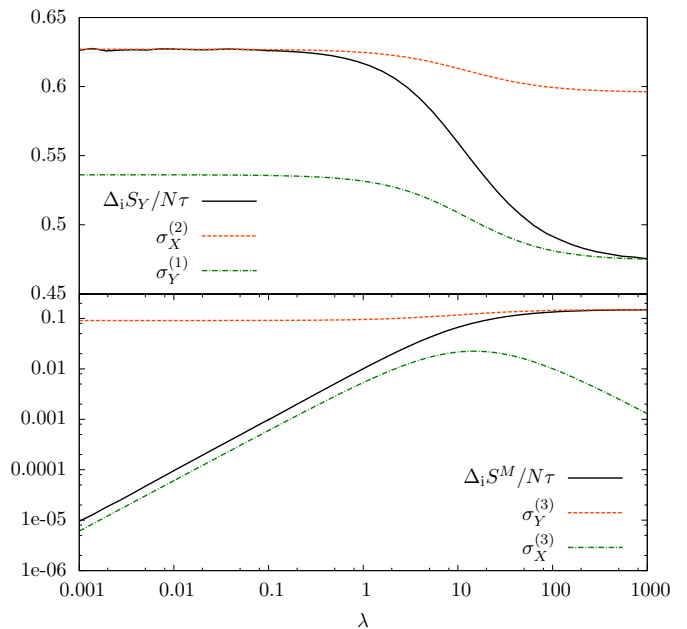}
\caption{\label{fig:SYSM}
Upper panel: Comparison between the numerical $\Di S_Y/N \tau$ (black, solid) and its asymptotic value $\sigma^{(2)}_{X}$ (red, dashed) and $\sigma^{(1)}_Y$ (green, dot-dashed) as a function of $\lambda$. 
Lower panel: Comparison between $\Di S^M/N \tau$ (black, solid), $\sigma^{(3)}_X$ (red, dashed) and $\sigma^{(3)}_Y$ (green, dot-dashed). 
Markovian discrete-time trajectories of length $N = 10^8$ and with time step $\tau= 10^{-4}$ have been considered. The set of parameters used is $a=10$, $r_{0}^+=0.15$, $r_{1}^+=0.1$, $r_{0}^-=0.1$, $r_1^-=0.2$, $w^{+}_{\rm on}= w^{+}_{\rm off}=0.1$, $w^{-}_{\rm on}=w^{-}_{\rm off}=1.3$, $k^+_{\rm off}=0.4$, $k^-_{\rm off}=0.3$.
}
\end{figure}

In Fig.~\ref{fig:SYSM}, we calculated numerically $\Di S^M/N \tau$ and $\Di S_Y/N \tau$ for this model by generating Markovian discrete-time trajectories in the joint space $(X,Y)$. Numerically $\Di S_X / N \tau$ is almost zero over the whole range in $\lambda$, therefore it is not shown. This is related to the fact that $\sigma_{X}^{(1)}$ is always zero for this model since states with different $x$ are connected by a single edge.
To interpolate between the two regimes of time-scale separation we introduced a scaling parameter $\lambda$ multiplying $r_0$ and $r_1$ and varying from $10^{-3}$ to $10^3$.\\ 
When $\lambda$ is small, $Y$ is faster than $X$ and the fast conditional dynamics of $Y$ at fixed state $x$ will reach a nonequilibrium steady-state obtained from (\ref{LocIntDyn}). As predicted by (\ref{eq:SYs2text}), $\Di S_Y / N \tau$ converges rapidly to $\sigma^{(2)}_X$. However, as explained below that equation, this convergence does not imply that the asymptotic value of the MEP will coincide with $\sigma^{(3)}_X$. The discrepancy between the two is related to the dependence of the fast rates on $X$, quantified in this model by the value of $a$ defined in {(\ref{SimpCond})}. We checked numerically that when $a$ approaches unity, this difference vanishes, consistently with our result in (\ref{eq:tssres}). Finally, the slow dynamics in $X$ becomes Markovian and, using (\ref{eq:slowX}), $\Di S_X / N \tau$ tends to $\sigma^{(1)}_{X}$ which is always zero in this model (not plotted).\\
We now turn to the opposite regime of time-scale separation at large values of $\lambda$, where $X$ is faster than $Y$. 
The fast conditional dynamics of $X$ at fixed state $y$ reaches a steady-state given by the solution of Eq.~(\ref{LocIntDyn}). In this case the steady-state corresponds to an equilibrium steady-state since transitions in $x$ are due to a single edge. Detailed balance is thus satisfied inside each state $y$. 
Under this condition we expect from (\ref{Ineq1-2}) that $\sigma^{(3)}_X=\sigma^{(2)}_Y={\mathcal{O}(\lambda^{-1})}$ (note that here the role of $X$ and $Y$ is exchanged compared to Sec.~{\ref{sec:TSsep}}). 
We also note that $\Di S_Y/N \tau$ approaches the Markovian rate $\sigma^{(1)}_Y$ as expected from (\ref{eq:slowX}) and that the MEP becomes exactly $\sigma^{(3)}_Y$ as predicted in Eq.~(\ref{eq:tssresBis}). Finally, according to (\ref{IntEqSY}), the entropy production $\Di S_X/N \tau$ remains very close to zero (not plotted).

\subsection{MEP as a measure of sensing}

We now consider two different models describing a sensing process. \\

\begin{figure}[h!]
\centering
\includegraphics{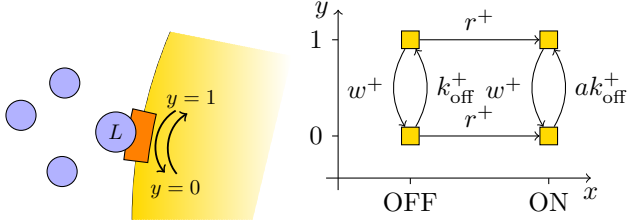}
\caption{\label{fig:modelI} 
ModelI: Simplified model of cellular signal transduction. The receptor $X$ transfers the information from the environment (presence/absence of the ligand molecule $L$) to the cell via the phosphorylation of an internal protein $Y$. The diagram shows the possible transitions between the four states of the model and the corresponding directed rates.
}
\end{figure}

{\bf Model I} is described in Fig.~(\ref{fig:modelI}) and represents an elementary model for biochemical signaling that was introduced in Ref.~\cite{SeifertBaratoPRE13}. 
A receptor $X$ switches with a rate $r^+$ ($r^-$) from (to) an inactive state $x={\rm off}$ to (from) an active state $x={\rm on}$ when binding (unbinding) its ligand molecule $L$ at fixed concentration in the environment. The receptor catalyzes the phosphorylation (dephosphorylation) of an internal protein $Y$ which will change as a result from (to) a dephosphorylated state $y=0$ to (from) a phosphorylated one $y=1$. This reaction depends on the receptor state $x$ and occurs at a rate $k_{x}^+$ ($k_{x}^-$). The factor $a$ in (\ref{SimpCond}) quantifies the relative catalytic activity of the receptor between its active and inactive state. The protein can also be dephosphorylated (phosphorylated) at a rate $w^+$ ($w^-$) by another enzyme which operates independently from the state of the receptor. This model is thus obtained from the model of section \ref{ModelGen} by assuming $w^{+/-}_{\rm on} = w^{+/-}_{\rm off} = w^{+/-}$ and $r_1^{+/-} = r_0^{+/-} = r^{+/-}$ so that the entropy production (\ref{EPAff}) becomes
\begin{align}
\sigma = (J_\mathrm{on}+J_\mathrm{off}) \ln \frac{k_{\rm on}^+ w^+}{k_{\rm on}^- w^-} . \label{EPAffBioMod}
\end{align}

\begin{figure}
\centering 
\vspace{.5cm}
\includegraphics[scale=.32]{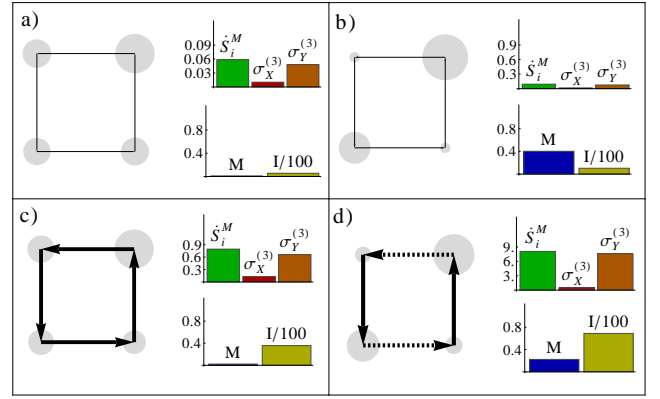}
\caption{\label{fig:grid} 
Model I in a regime of a) low flux circulation and low mutual information b) low flux and high mutual information c) high flux circulation and low mutual information d) high flux circulation and high mutual information. This latter correspond to a regime of time-scale separation (slow transitions are dashed). The MEP rate $\bull{S}^M_{\rm i}(0)$ (\ref{eq:mep0Ineql}), $\sigma^{(3)}_X$ and $\sigma^{(3)}_Y$ (\ref{eq:S3}), the mutual information $M$ (\ref{eq:Mxy}), and the mutual information rate $\bull{I}(0)$ (\ref{eq:ub}) are displayed in the different regimes.
}
\end{figure}

This system operates as an accurate sensor for the cell when the state of the protein $Y$ rapidly responds to the detection of a ligand molecule by the receptor and correlates its state to the receptor state $x$. To be accurate, the protein dynamics has to be fast compared to the time scale of the ligand biding and unbinding. In other words, when the receptor gets activated (resp. deactivated), the protein needs to respond rapidly and causally by becoming phosphorylated (resp. dephosphorylated). For such an effect to occur, not only the mutual information (\ref{eq:Mxy}) needs to be significant at steady-state, but also an important counterclockwise flux circulation along the large cycle on Fig.~(\ref{fig:modelI}) is required. 
This latter enables the fast causal response of the protein state. The magnitude of this circulation (i.e. the net probability flux along the large cycle) is displayed in Fig.~(\ref{fig:grid}) for different regimes of model I. We note that a high MEP as well as a high $\sigma^{(3)}_Y$ are obtained when both the mutual information and the flux circulation are high, i.e. precisely in the regime where accurate signaling occurs. In this regime, since the transitions in $x$ do not depend on $y$, using (\ref{eq:tssres}) we find that $\bull{S}^M_\mathrm{i}=\sigma^{(3)}_Y$.
We note that under physiological conditions, the flux circulation along the two small cycles should also occur in the counterclockwise direction. This specific condition is typically reached using $k_{\rm on}^+ \gg k_{\rm on}^-$ and $w^{+} \gg w^{-}$.\\

\begin{figure}[h]
\includegraphics{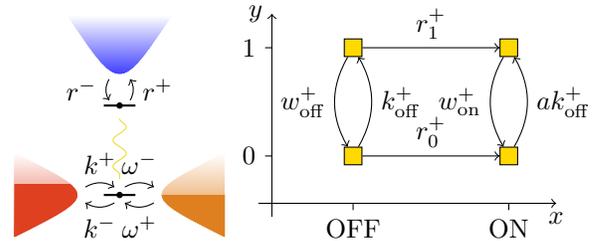}
\caption{\label{fig:modelII} 
Model II: The upper single level quantum dot $X$ (in contact with a cold lead) is sensing via capacitive coupling the presence or absence of an electron in the lower dot $Y$ (in contact with two warmer leads). The diagram shows the possible transitions between the four states of the model and their corresponding directed rates.
}
\end{figure}

{\bf Model II} is made of two capacitively coupled single level quantum dots as depicted on Fig. \ref{fig:modelII}. 
\begin{figure}[h]
\centering 
\vspace{.5cm}
\includegraphics[scale=.35]{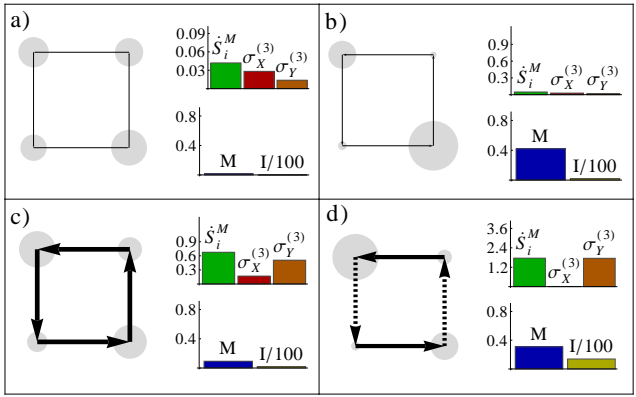}
\caption{\label{fig:grid_MD} 
Same as Fig. (\ref{fig:grid}) but for Model II. 
}
\end{figure}
It is defined from the general model of section \ref{ModelGen} by assuming that the second and third affinities in (\ref{EPAff}) are the same, namely under the condition
\begin{align}
w^{-}_{\rm on} w^{+}_{\rm off} = w^{+}_{\rm on} w^{-}_{\rm off}.
\end{align}
Its entropy production is therefore of the form
\begin{align}
\sigma = J_\mathrm{c} \ln \frac{r_0^+ r_1^- }{r_0^- r_1^+} 
+ (J_\mathrm{on}+J_\mathrm{off}) \ln \frac{k_{\rm on}^+ w^+_{\rm on}}{k_{\rm on}^- w^-_{\rm on}} . \label{EPAffMaxDemMod}
\end{align}
In this model $Y$ is a single level quantum dot in contact with two leads at same temperature but different chemical potentials, while $X$ is a second single level quantum dot capacitively coupled to the first dot and in contact with a lead at lower temperature. Such models have been used to describe single electron detectors in electron counting statistics \cite{BrandesSchallerPRB10, EspositoCuetaraGaspard11, ButtikerSanchez10PRL}. 
In Ref.~\cite{EspositoStrassSchall12} this model has been used to show that in a finely tuned regime, dot $X$ could play the role of an ideal Maxwell demon acting on dot $Y$. 

In this paper we focus on a broader regime where $X$ can accurately sense the electron transfers in and out of $Y$, by causally correlating its state with the state of $Y$. 
Typically, if an electron enters (exits) $Y$ from one of its two leads, the state of $X$ has to immediately become empty (filled). This implies that the dynamics of $X$ has to be much faster with respect to $Y$ and also that the mutual information between $X$ and $Y$ has to be large as well as the counterclockwise probability flux along the large cycle, in order to generate the causal response of $X$. This probability flux is displayed in Fig.~(\ref{fig:grid_MD}) for different regimes of model II. We observe that the combination of high MEP and high $\sigma^{(3)}_Y$ corresponds to the regime of accurate sensing. Since in this regime the conditional dynamics in $X$ at fixed $y$ equilibrates, we have $\sigma_{Y}^{(2)}=0$, thus, using (\ref{eq:mep0Ineql}), (\ref{Ineq1-2}) and (\ref{eq:tssresBis}), we again find that $\bull{S}^M_\mathrm{i}=\sigma^{(3)}_Y$. 
Finally we observe that in the ideal Maxwell demon regime, $\sigma_{Y}^{(1)}$ can be interpreted as the entropy production generated by a Markov dynamics in $Y$ with rates phenomenologically modified as proposed in \cite{EspositoSchaller12} to account for a Maxwell demon feedback \cite{EspositoStrassSchall12}. 
However, such a phenomenological approach neglects $\sigma_{Y}^{(3)}$, which is by far the dominant contribution to the total entropy production of the process and which diverges in the regime of perfect detection.\\

Despite significant differences between the two sensing models proposed in this section, we found in both cases that the MEP is given by $\bull{S}^M_\mathrm{i}=\sigma_{Y}^{(3)}$. These results suggest that the MEP could provide a meaningful thermodynamic measure for the detection accuracy. 

\section{Conclusions}\label{conc}

We introduced in this paper the notion of mutual entropy-production (MEP) and showed that it can be expressed as the difference between the mutual information rate and the time-reversed mutual information rate. 
This result is analogous to the expression of the entropy production as the difference between the time-reversed entropy rate and the entropy rate, as found by Gaspard in \cite{Gaspard04b}. 
The MEP is in general hard to evaluate due to the non-Markovian character induced by coarse-graining procedures. 
However, for a bipartite system we were able to provide explicit expressions in the short-time limit and in the presence of time-scale separation between its components.
We also verified numerically the accuracy of these results in a four-state model system. 
Based on the study of two simple but very different models of detection, one used in \cite{SeifertBaratoPRE13} to describe biochemical signal transduction and the other used in \cite{EspositoStrassSchall12} to describe single electron detection, we provided evidence that the MEP could be a relevant thermodynamic measure for sensing in several frameworks.

\section{Acknowledgments}

This work is supported by the National Research Fund, Luxembourg in the frame of project FNR/A11/02.

\end{document}